\documentclass{article}
\usepackage{graphicx} 
\usepackage{amsmath}
\usepackage{amsfonts}
\usepackage{amssymb}
\usepackage{amsbsy}
\usepackage{mathrsfs}
\usepackage{amsthm}
\usepackage{graphicx}    
\usepackage{rotating}    
\usepackage{epsfig}
\usepackage{ulem}
\usepackage{color}       
\usepackage{xcolor}

\definecolor{gruen}{rgb}{0,0.625,0}     
\definecolor{rot}{rgb}{0.75,0,0}        
\definecolor{blau}{rgb}{0,0,0.75}       

\newcommand{\BEQ}{\begin{equation}}     
\newcommand{\BEA}{\begin{eqnarray}}
\newcommand{\BD}{\begin{displaymath}}
\newcommand{\EEQ}{\end{equation}}       
\newcommand{\EEA}{\end{eqnarray}}
\newcommand{\ED}{\end{displaymath}}
\newcommand{\D}{{\rm d}}                
\newcommand{\II}{{\rm i}}               
\renewcommand{\Im}{{\rm Im\ }}          
\newcommand{\demi}{\frac{1}{2}}         

\newcommand{\appsection}[2]{\setcounter{equation}{0}\setcounter{subsection}{0}\setcounter{table}{0}\setcounter{figure}{0}
\section*{Appendix #1. #2}
\renewcommand{\theequation}{#1.\arabic{equation}}
              \renewcommand{\thesection}{#1}\renewcommand{\thetable}{#1\arabic{table}}\renewcommand{\thefigure}{#1.\arabic{figure}} }

\begin{document}

\begin{titlepage}

\vskip 1.5 cm
\begin{center}
{\Large \bf Fractional cosmic strings}
\end{center}

\vskip 2.0 cm
\centerline{{\bf S\'ebastien Fumeron}$^{a}$, {\bf Malte Henkel}$^{a,b}$ and {\bf Alexander Lop\'ez}$^{c}$}
\vskip 0.5 cm
\begin{center}
$^a$Laboratoire de Physique et Chimie Th\'eoriques (CNRS UMR 7019),\\  Universit\'e de Lorraine Nancy,
B.P. 70239, F -- 54506 Vand{\oe}uvre l\`es Nancy Cedex, France\\
$^b$Centro de F\'{i}sica Te\'{o}rica e Computacional, Universidade de Lisboa, \\Campo Grande, P--1749-016 Lisboa, Portugal\\
$^c$Escuela Superior Polit\'ecnica del Litoral, ESPOL,\\ Departamento de F\'isica, Facultad de Ciencias Naturales y Matem\'aticas,\\ Campus Gustavo Galindo
 Km. 30.5 V\'ia Perimetral, P. O. Box 09-01-5863, Guayaquil, Ecuador \\~\\
\end{center}

\begin{abstract}

Topological defects in the framework of effective quantum gravity model are investigated, based on the
hypothesis of an effective fractal dimension of the universe. This is done by using Caputo fractional derivatives to determine the spacetime geometry of a fractional cosmic string. Several results for the propagation of light are discussed, notably the light deviation angle due to the defect and the geodesics of light. \\[0.9cm]~
\centerline{\large \today}
\end{abstract}

\vfill

\end{titlepage}

\setcounter{footnote}{0}

\section{Introduction}


One of the most exciting challenges in modern cosmology is the quest for a quantum theory of gravity at the Planck scale. Apart from the two major contenders, 
namely the fully quantized super-string theory and quantum loop gravity, there is a growing interest in the search for effective theories of gravity 
that only include quantum corrections, but have the asset of being far more convenient. 

Surprisingly, although built upon different premises, effective field theories seem to conspire towards a dependence of the dimensionality with the energy scale: models such as causal dynamical triangulations 
\cite{ambjorn2005spectral}, asymptotically safe gravity \cite{lauscher2005fractal}, spin-foams \cite{modesto2009fractal}, $\kappa$-Minkowski spacetimes \cite{benedetti2009fractal} 
or Ho{\v{r}}ava-Lifshitz gravity \cite{hovrava2009spectral} agree upon the existence of an emerging time-space with an effective 
dimension\footnote{The definition used here is the spectral dimension built from the first derivative of the normalised trace of the heat kernel (see for instance Eq.~(2) in \cite{benedetti2009fractal}).} 
$d_s=4$ at large scales, which reduces down to $d_s=2$ when approaching the Planck scale. Such scale-dependent behaviour suggests time-space properties might be fractal \cite{calcagni2021multifractional}, 
a feature that was also foreseen a long time ago by Nottale \cite{nottale1993fractal}. 

In that perspective, fractional calculus seems a convenient tool to investigate scale-dependent gravity \cite{calcagni2021classical}. 
Yet, up to now, only few studies have implemented fractional calculus in cosmology. In \cite{el2013fractional}, 
the fractional Einstein equations were derived from the fractional action principle by using the modified Riemann-Liouville operator: 
application to the gravitational relaxation of Friedman-Lema\^{\i}tre-Robertson-Walker (FLRW) time-space shows discrepancies with Einstein’s general relativity 
that could account for the current accelerated expansion of the universe. In \cite{garcia2022cosmology}, 
the authors used a fractional action principle (based on the Caputo fractional derivative) built from a FLRW ansatz to derive the fractional Friedman equations: 
they set stringent constraints on the fractional parameter to fit to observational data. In another remarkable work \cite{di2023vacuum}, 
the authors determined a vacuum solution for a static and spherically-symmetric time-space based on the fractional Riemann-Liouville derivative: 
while the exterior Schwarzschild's metric is retrieved outside the horizon, the interior solutions are not only regular at $r=0$, but depending on the fractional parameter, they can match with gravastar models. 

In this paper, we apply fractional calculus to determine the fractional version of an extensively studied object: the Nambu-Goto cosmic string. In the next section, we recall the basics of these topological defects (the basic ansatz, form of Einstein equations, geometrical properties) to prepare for the last-step modification method. 
In section 3, we use the Caputo derivative to obtain the metric corresponding to a fractional cosmic string and determine some of its main gravitational effects on light propagation 
in sections 4 and 5, namely deviation of light induced by the defect and the geodesics of light. Our conclusions are given in section~6. Two appendices provide technical background. 

\section{Nambu-Goto strings in a nutshell}

Cosmic strings are generic relics of the phase transitions that occurred in the early universe \cite{jeannerot2003generic}. 
The simplest cosmic defects one may expect in cosmology are the Nambu-Goto strings, which consist of delta-distributed linear concentrations of mass-energy $\mu$. 
In the Landau-Lifshitz sign convention and setting $c=1$, the metric of such objects can be assumed as 
\begin{equation}
\D s^2=e^{A(r)}\left(\D t^2-\D z^2\right)-\D r^2-e^{B(r)}\D \theta^2 \label{NG}
\end{equation}
Solving Einstein's equations in the vacuum (where $T_{\mu\nu}=0$), one finds the system \cite{vilenkin1994}:
\begin{subequations}
\begin{align}
A''+A'^2+\frac{1}{2}A'B'&= 0 \label{e1} \\
B''+\frac{1}{2}B'^2+A'B'&= 0 \label{e2} \\
2A''+B''+A'^2+\frac{1}{2}B'^2&= 0 \label{e3}
\end{align}
\end{subequations}
with $A'(r)=\frac{\D A(r)}{\D r}$. Linear combination of these three equations leads to
\begin{eqnarray}
A'\left(A'+2B'\right)=0 \label{e4}
\end{eqnarray}
The physically relevant solution is the one corresponding to $A'(r)=0$ \cite{vilenkin1994}, which gives $A(r)=a$ and using eq.~(\ref{e2}), $B(r)=2\ln(r)+b$. The ansatz (\ref{NG}) then becomes  
\begin{equation}
\D s^2=e^a\left(\D t^2-\D z^2\right)-\D r^2-e^b r^2\D \theta^2 \label{NG2}
\end{equation}
After re-scaling the radial coordinate, one obtains (up to a constant conformal factor)
\begin{equation}
\D s^2=\D t^2-\D r^2-\alpha^2r^2 \D\theta^2-\D z^2 \label{Vilenkin2}
\end{equation}
This is the well-known Vilenkin metric \cite{vilenkin1981gravitational} describing time-space in the presence of a cosmic string. Physically the defect parameter is related to the string mass-energy density 
$\mu$ by $\alpha=1-4G\mu<1$. The time-space geometry described by (\ref{Vilenkin2}) is conical and can be thought of as the result of a Volterra cut-and-glue process, 
the angular sector removed to generate the cone being $2\pi(1-\alpha)$ \cite{fumeron2023introduction}. 

Data collected by Planck collaboration set stringent limits on the the string parameter $\mu$ \cite{ade2014planck} and for a GUT-scale string, this latter is estimated at about $10^{16}$ tons per meter: in the Volterra picture, this corresponds to a deficit angle of a few seconds of arc. Such geometry is likely to leave several observational imprints in the form of planar wakes and their impact on the filamentary structure of the cosmic web \cite{fernandez2020cosmic}, or temperature discontinuities in the cosmic microwave background \cite{kaiser1984microwave}. Although the existence of these defects is still not settled, a recent work suggested that the North American Nanohertz Observatory for Gravitational Waves could have found evidence for cosmic strings in the stochastic gravitational wave background \cite{blasi2021has}.

Before performing the fractional calculus analysis, we rewrite equation (\ref{e2}) in terms of the function $F(r)=e^{B(r)/2}$, such that we get
\begin{equation}\label{simpler}
F''(r)=0
\end{equation}
which integrates to the deficit angle  as $F'(r)=\alpha$. This result will be used in the next section to discuss the fractional relativity.

\section{Metric of a fractional string}

We now switch to the fractional counterpart of Nambu-Goto strings. 
There are several ways to define a well-behaved fractional derivative operator. Throughout this paper, we consider the Caputo fractional derivative of order $\eta$ (see appendix~B):
\begin{equation}
^c D^{\eta}\: f(x)=\frac{1}{\Gamma(n-\eta)}\int_{e}^x \!\D u\: \frac{1}{(x-u)^{\eta-n+1}}\frac{\D^n f(u)}{\D u^n}
\end{equation} 
where $n=\eta$ if $\eta\in\mathbb{N}$ and $n=\left[\eta\right]+1$ otherwise. The main asset of this definition is that the Caputo derivative of a constant vanishes, as for ordinary derivatives. For the rest of the paper we assume, without loss of generality, that $e=0$ and $0\leq \eta<1$.

Fractional equations are obtained by performing the last step modification (or LSM) procedure \cite{di2023vacuum}: 
it consists in replacing the ordinary derivatives with their fractional counterparts to avoid combinatoric issues related to the Leibniz rule. Using this prescription in 
(\ref{e4}), one finds that
\begin{eqnarray}
^c D^{\eta}A=0
\end{eqnarray}
which still means that $A(r)=a$ is a constant, but when it comes to (\ref{simpler}), one now has
\begin{equation}
^{c}D^{\eta}F_{\eta}(r)=\alpha.
\end{equation}
Using the definition of the Caputo derivative of power functions \cite{herrmann2011fractional}
\begin{equation}
^c D^{\eta}x^k=\frac{\Gamma(1+k)}{\Gamma(1+k-\eta)}x^{k-\eta}\qquad;\quad x\geq0,\quad k\neq-1,-2,-3\dots
\end{equation}
this leads to
\begin{equation}
F_\eta(r)=\frac{\alpha}{\Gamma(1+\eta)}r^\eta.
\end{equation}
With this solution, the metric would read as
\begin{equation}
\D s^2 =\D t^2-\D z^2 -\D r^2-\left(\frac{\alpha}{\Gamma(1+\eta)}\right)^2 r^{2\eta} \D\theta^2 \label{eta-string}
\end{equation}
In the limit $\eta\rightarrow 1$, the Vilenkin metric (\ref{Vilenkin2}) is recovered. As the time component is trivial ($g_{00}=1$), there is no gravitational pulling exerted by a fractional string onto neighbouring objects.

The geometry associated with the metric (\ref{eta-string}) is not that of an ordinary cone. Yet, there is a difference between the angular circumference in flat time-space and in the field of the fractional string: the radial distance being identical to the circumferential radius ($g_{rr}=1$), the angular mismatch is 
\begin{equation}
\delta(r)=2\pi\left(1-\frac{1}{\sqrt{g_{rr}}}\frac{\D}{\D r}\sqrt{g_{\theta\theta}}\right)=2\pi\left(1-\frac{\alpha \eta }{\Gamma(1+\eta)} r^{\eta-1}\right).
\end{equation}
Contrary to the Nambu-Goto string, the angular mismatch is not a positive constant (conical geometry) and it depends not only on the string mass-energy density but also on the fractional parameter $\eta$. Besides, $\delta(r)$ undergoes a sign change at $\bar{r}=\left[\Gamma(1+\eta)/(\alpha\eta)\right]^{\frac{1}{\eta-1}}$: if $\eta<1$, then for $r<\bar{r}$, $\delta(r)<0$ and the geometry is saddle-like, whereas for $r>\bar{r}$, $\delta(r)>0$ and the geometry is that of a flaring cone\footnote{The Caputo derivative holds only for $0\leq\eta<1$, so the case where $\eta>1$ is dismissed in this work.}. This implies that for a moving fractional string, expected large-scale kinematic signatures such as planar wakes or the Kaiser-Stebbins effect should be strongly reconsidered.

The Killing vectors associated to (\ref{eta-string}) are $(K_t)^{\mu}=(1,0,0,0)$, $(K_{\theta})^{\mu}=(0,0,1,0)$ and $(K_z)^{\mu}=(0,0,0,1)$. This provides us with three first integrals
\begin{subequations} \label{conserve}
\begin{align}
&(K_t)_{\mu}\frac{\D x^{\mu}}{\D\lambda}=\frac{\D t}{\D \lambda}=E &&\text{(energy)} \label{consE} \\
&(K_{\theta})_{\mu}\frac{\D x^{\mu}}{\D\lambda}=\left(\frac{\alpha r^\eta}{\Gamma(1+\eta)}\right)^2\frac{\D\theta}{\D\lambda}=L &&\text{(angular momentum)}  \label{consL} \\
&(K_z)_{\mu}\frac{\D x^{\mu}}{\D\lambda}=\frac{\D z}{\D\lambda}=P_z &&\text{(momentum along $z$)}  \label{consP}
\end{align}
\end{subequations}
where $\lambda$ parametrises a geodesic path. For massive particle, the parametrisation is taken as $\lambda=\tau$, with $\tau$ the proper time. 
Considering motions confined in planes orthogonal to the defect axis, substituting (\ref{conserve}) 
into (\ref{eta-string}) gives the radial equation
\begin{equation} \label{eq15}
\frac{1}{2}(E^2-1)=\tilde{E}=\frac{1}{2}\left(\frac{\D r}{\D\tau}\right)^2+\frac{\Gamma(1+\eta)^2 L^2}{2\alpha^2}\frac{1}{r^{2\eta}}\geq 0
\end{equation}
This is an energy conservation equation per unit mass, the last term being a purely repulsive potential energy. 
The case of a particle with a non-vanishing momentum along \textit{z} simply changes the value of the constant $\tilde{E}$ but not the qualitative effect of the potential. 
Hence as for Nambu-Goto strings, no bound state can be expected in the gravitational field of a fractional cosmic string. 

\section{Lensing effects}

In some cosmological models, topological defects, in the form of cosmic strings, are predicted to generate gravitational lensing effects, 
via a deficit angle that is proportional to the string's density mass  \cite{vilenkin1985cosmic}. 
Within the standard analysis, this topological feature is found to be independent of the impact parameter, 
associated to the deflection of light rays.
\begin{figure}[tb]
\includegraphics[height=2.5in]{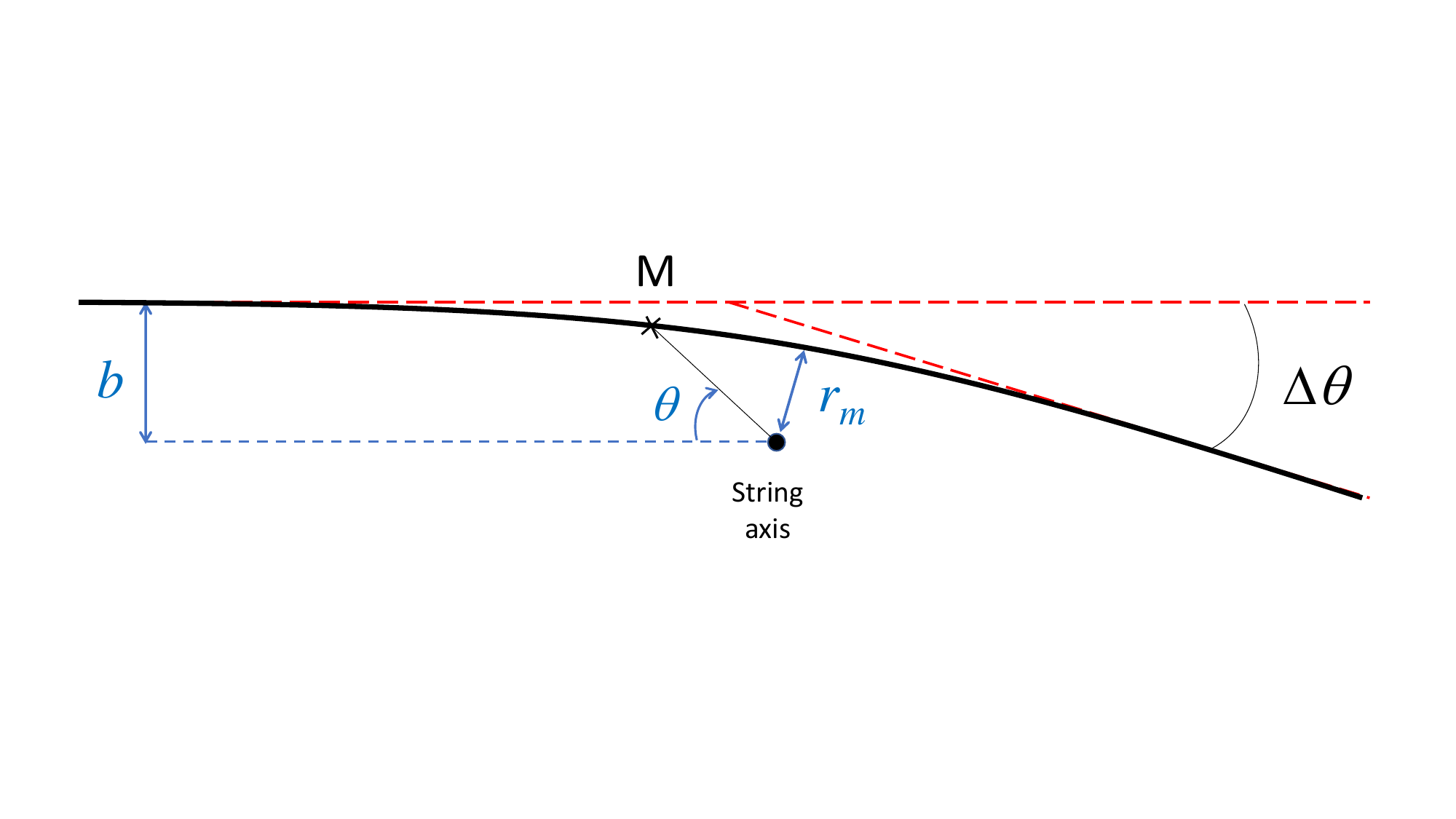} 
\caption{\textit{Deviation of a light ray (black thick line) in the presence of a fractional cosmic string.}}\label{fig1}
\end{figure}
For light, one defines the impact parameter as (see figure~\ref{fig1})
\begin{equation}
b=\frac{L}{E}=\left(\frac{\alpha r^\eta}{\Gamma(1+\eta)}\right)^2\frac{\D\theta}{\D t}
\end{equation}
Light-like geodesics are obtained when $\D s^2=0$. The deviation of light rays due to the string is determined from the formula
\begin{eqnarray}
\Delta \theta&=&2\int_{r_m}^{+\infty}\frac{\D\theta}{\D r}\D r-\pi=2\int_{r_m}^{+\infty}\left(\sqrt{\frac{-g_{rr}g_{tt}b^2}{g_{\theta\theta}g_{tt}b^2+g_{\theta\theta}^2}}\right) \D r-\pi \nonumber \\
&=&\frac{R^2}{b}\int_{r_m}^{+\infty}\frac{\D r}{r^{\eta}\sqrt{r^{2\eta}-R^2}}-\pi
\end{eqnarray}
with $R^2=b^2\Gamma(1+\eta)^2/\alpha^2$. The distance of closest approach $r_m=R^{1/\eta}$ is obtained from the condition  
$\left.\frac{\D r}{\D\theta}\right|_{\theta=\theta_m}=0$ and $r_m=r(\theta_m)$. 
Hence the angular deviation becomes 
\begin{eqnarray}
\Delta \theta&=&2\int_{r_m}^{+\infty}\frac{\D\theta}{\D r}\D r-\pi \:=\: 2\int_{r_m}^{+\infty}\left(\sqrt{\frac{-g_{rr}g_{tt}b^2}{g_{\theta\theta}g_{tt}b^2+g_{\theta\theta}^2}}\right) \D r-\pi 
\nonumber \\
&=&2\frac{r_m^{2\eta}}{b}\int_{r_m}^{+\infty}\frac{\D r}{r^{\eta}\sqrt{r^{2\eta}-r_m^{2\eta}}}-\pi \nonumber \\
&=&2\frac{r_m}{b}\int_{1}^{+\infty}\frac{\D u}{u^{\eta}\sqrt{u^{2\eta}-1}}-\pi 
\:=\: \frac{r_m}{b}\frac{\sqrt{\pi}}{\eta}\frac{\Gamma(1-1/2\eta)}{\Gamma(3/2-1/2\eta)}-\pi
\label{eq25}
\end{eqnarray}
where we used the identity, for $\eta>\demi$ 
\begin{equation} \label{eq26}
I_\eta :=\int_{1}^{+\infty}\frac{\D u}{u^{\eta}\sqrt{u^{2\eta}-1}} 
=\frac{\sqrt{\pi}}{2\eta}\frac{\Gamma(1-1/2\eta)}{\Gamma(3/2-1/2\eta)} 
\end{equation}
Interestingly, the validity of this result enforces the condition $\eta>1/2$ which tells us that $\eta=1/2$ 
needs special consideration. Besides, in the case where $\eta\rightarrow 1$, one recovers the standard result
\begin{equation}
\Delta\theta =\frac{\pi}{\alpha}-\pi.
\end{equation}
Ought to the cylindrical symmetry, the lensing effect due to a Nambu-Goto string gives rise to double images of a far-away source, on each side of the string, 
in contrast with Einstein rings which are expected in the case of spherically symmetric lens. 
In the case of fractional cosmic strings, one may expect the stretching of the double images as the deviation angle depends on the impact parameter in the form $b^{1/\eta-1}$.

\section{Light geodesics}

\begin{figure}[t]
\includegraphics[width=2.25in]{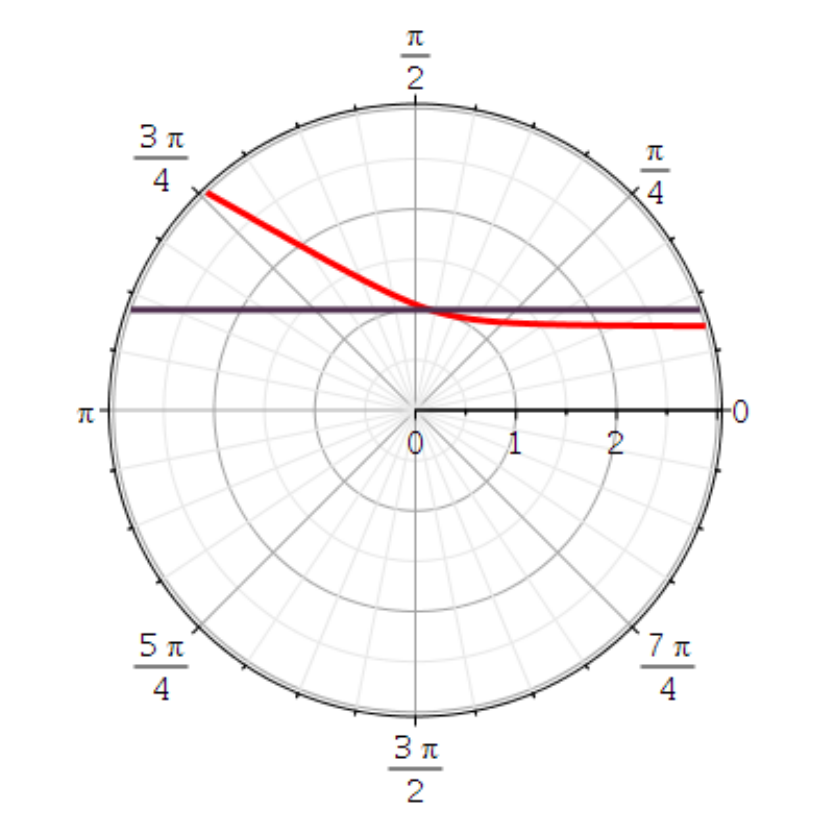} ~\includegraphics[width=2.25in]{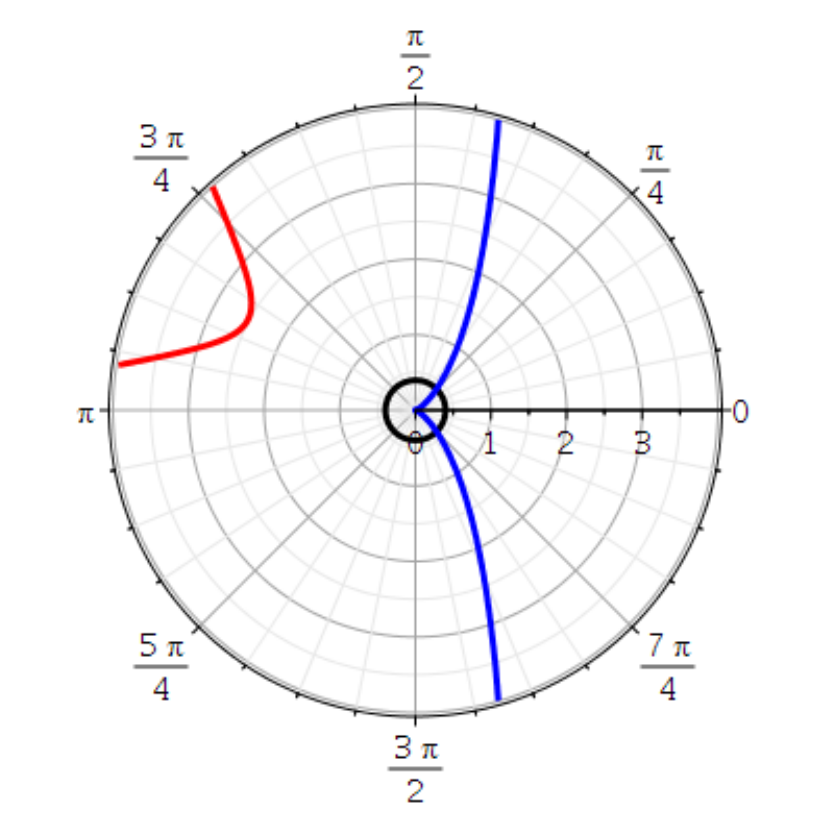} ~\includegraphics[width=2.25in]{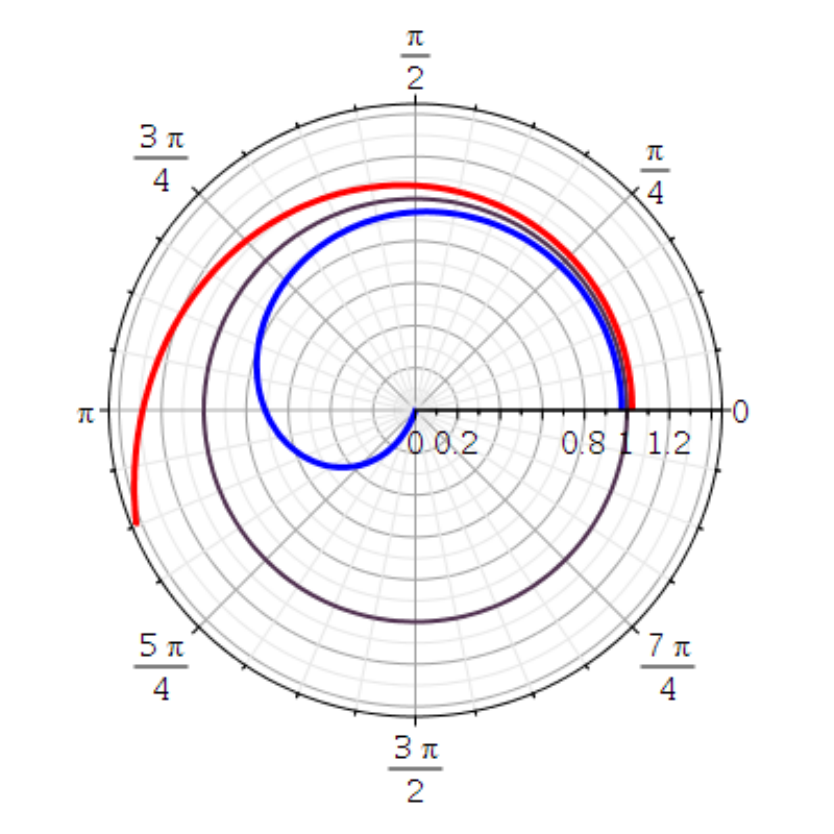} 
\caption[fig4]{\it Geodesics $r=r(\theta)=u(\theta)^{1/(1-2\eta)}$ of light in polar coordinates. \\
\underline{Left panel:} $\eta=1$. Geodesics for $\alpha=1$ (violet) and for $\alpha=1.2$ (red) are shown. We used $\alpha=b$.\\
\underline{Middle panel:} $\eta=3/4$. There is a constant-radius geodesic $r=r_0$ (black), and two geodesics which escape to infinity (red \& blue). 
We used ${\cal A}={\cal B}=1$ and $\theta_0=0$.\\
\underline{Right panel:} $\eta=1/2$. Geodesics are shown for $r(0)<r_c$  (blue curve) and $r(0)>r_c$ (red curve) 
and $\frac{\D r}{\D \theta}(0)=0$. The limit cycle at $r=r_c=1$ is also indicated (see main text).}
\label{fig4}
\end{figure}

In analogy to (\ref{eq15}), null geodesics are derived by solving the equation 
\begin{equation}
\left(\frac{\D r}{\D t}\right)^2+\frac{\Gamma(1+\eta)^2 b^2}{\alpha^2 r^{2\eta}}=1
\end{equation}
Equivalently, the geodesic orbit $r=r(\theta)$ for light is given by 
\begin{equation} \label{geo-lumiere}
\left(\frac{\D r}{\D\theta}\right)^2+\left(\frac{\alpha r^\eta}{\Gamma(1+\eta)}\right)^2=\frac{1}{b^2}\left(\frac{\alpha r^\eta}{\Gamma(1+\eta)}\right)^4
\end{equation}
This last equation (\ref{geo-lumiere}) can be reduced to a simpler form through the ansatz $r=u^{1/\beta}$, where $\beta$ is found from 
\BEQ \label{cond:etabeta}
\frac{4\eta}{\beta} - \frac{2}{\beta} +2 \stackrel{!}{=} 0
\EEQ
which eliminates the strongest non-linearity. It follows that $\beta=1-2\eta$ and 
\BEQ \label{eq42}
\left( \frac{\D u}{\D\theta}\right)^2 +  \left( \frac{\alpha (1-2\eta)}{\Gamma(1+\eta)}\right)^2 
u^{2\eta/(2\eta-1)} = \frac{(1-2\eta)^2}{b^2} \frac{\alpha^4}{\Gamma^4(1+\eta)} \;\; ; \;\; 
\eta\ne \demi
\EEQ
\noindent 
such that for $\eta=1$ this reduces to the equation $\left(\frac{\D u}{\D\theta}\right)^2 + \alpha^2 u^2 = \frac{\alpha^4}{b^2}$. 
Clearly, for $\alpha=1$ this reproduces a straight line in polar coordinates, as expected.
However, for $\alpha\ne 1$ a curved geodesic is obtained, see the left panel of figure~\ref{fig4}. This illustrates how straight light rays are bent through a cosmic string. 

In general, a reduction of (\ref{eq42}) to integrals is
straightforward but will involve ill-known special functions. 
\noindent In the simplest case, a reduction of the solutions to elliptic functions is possible. 
For example, if $\eta=\frac{3}{4}$, from (\ref{eq42}) the geodesic curves are solution of the equation
\BEQ \label{eq41}
\left( \frac{\D u}{\D\theta}\right)^2 + 4{\cal A} u^3 = {\cal B} \;\; ; \;\; 
\eta = \frac{3}{4}
\EEQ
with constants ${\cal A},{\cal B}>0$.
Eq.~(\ref{eq41}) is either solved by the constant $u(\theta)= \left( \frac{\cal B}{4{\cal A}}\right)^{1/3}$ 
or else by a Weierstra{\ss} $\wp$ function (see appendix~A for details)
\BEA
u(\theta) &=& - \frac{1}{{\cal A}^{1/3}}\: \wp\left( {\cal A}^{1/3} \bigl( \theta + \theta_0\bigr); 0, -{\cal B}\right) 
           = \frac{1}{{\cal A}^{1/3}}\: \wp\left( \II {\cal A}^{1/3} \bigl(\theta + \theta_0\bigr); 0, +{\cal B}\right) 
\nonumber \\
          &=&  \left(\frac{\cal B}{\cal A}\right)^{1/3} \wp\left( \II ({\cal A}^2 {\cal B})^{1/6}\bigl(\theta + \theta_0\bigr); 0, +1\right)
\EEA 
In the middle panel of figure~\ref{fig4}, their geometric form in the $(r,\theta)$-plane is illustrated and should be qualitatively similar for all 
$\demi < \eta <1$. Besides the circle, one class of geodesics $r=r(\theta)=1/u(\theta)^2$ passes through the
origin which a characteristic cusp (blue curve) and the other class arrives from infinity, passes at the same minimal distance from the origin, and escapes to infinity (red curve). 
For $\alpha=1$, both geodesics should become straight lines in the $\eta\to 1$ limit and comparison with figure~\ref{fig4} shows how the fractional nature of the parameter $\eta$ impacts on the paths of light rays.   

The case $\eta=\demi$ is analysed separately. From (\ref{geo-lumiere})
\BEQ
\left( \frac{\D r}{\D\theta}\right)^2 + \left( \frac{\alpha}{\Gamma(3/2)}\right)^2 r = 
\frac{\alpha^4}{b^2 \Gamma^4(3/2)} r^2 \;\; ; \;\; \eta=\demi 
\EEQ
Taking a further standard derivative with respect to $\theta$, if $\frac{\D r}{\D\theta}\ne 0$ the orbit follows from a Binet-type formula 
\BEQ
\frac{\D^2 r}{\D \theta^2} = \frac{\alpha^4}{b^2 \Gamma^4(3/2)} r - \demi \left( \frac{\alpha}{\Gamma(3/2)}\right)^2 = 
\frac{\alpha^4}{b^2 \Gamma^4(3/2)} \left[ r - \frac{b^2\Gamma^2(3/2)}{2\alpha^2} \right] \;\; ; \;\; \eta=\demi 
\EEQ
First of all, see right panel of figure~\ref{fig4}, we see a limit cycle of radius $r_c = \frac{b^2\Gamma^2(3/2)}{2\alpha^2}>0$. For radii
such that $r<r_c$, there is an inward spiral and after a rotation on an angular scale of 
$\omega^2 = \left(\frac{2\pi}{\theta_{\rm osc}}\right)^2 = \frac{\alpha^4}{b^2 \Gamma^4(3/2)}$, the radius has shrunk to zero. In other words, in this regime, light cannot escape the gravitational field of the string and gets trapped at the defect core.
On the other hand, for $r>r_c$, the generic motion is away from the limit cycle, and only for very special initial conditions  one might find a convergence to it.  
The geodesics shown in the right panel of figure~\ref{fig4} for $\eta=\demi$ correspond to half the trajectories shown in the middle panel of figure~\ref{fig4}, for $\eta=\frac{3}{4}$. 
Important differences with respect to the spiralling around the origin can be seen.


\section{Concluding remarks}

In this work, we determined the metric surrounding a cosmic string in the context of fractional cosmology. Our calculation, based on the Caputo derivative, illustrates how the corresponding time-space geometry depends both on the mass-energy density of the string and on the fractional dimension $\eta$. 

Instead of double point-like images of far-away sources, the lensing effects due to a fractional cosmic string are likely to produce stretched double images as the deviation angle depends on the impact parameter in the form $b^{1/\eta-1}$. The light paths are strongly dependent on the fractional parameter, with the case $\eta=1/2$ giving rise to possible trapping effects towards the defect core. 

Because the time-space geometry of such defect is no longer that of a cone, one may expect other significant observational signatures. One of them is the Kaiser-Stebbins effect: for a moving Nambu-Goto string, this mechanism predicts the existence of step-like discontinuities in the cosmic microwave background temperature \cite{kaiser1984microwave}. Up to now, it has not been detected for regular string and a possible extension of the present work could consist in implementing the Kaiser-Stebbins effect for a fractional defect. This will be treated in a forthcoming paper.



\appsection{A}{On the Weierstra{\ss} $\wp$-function}

Some background is provided on the elliptic $\wp$-functions,  needed in solving (\ref{eq42}) and which is taken from \cite{Abra65}. 

An elliptic function has two complex half-periods $\omega,\omega'\in\mathbb{C}$ with $\Im \frac{\omega'}{\omega}>0$ and 
such that $f(z)=f(z+2M\omega + 2 N\omega')$ for $N,M\in\mathbb{Z}$ and $z\in\mathbb{C}$. The special Weierstra{\ss} function $\wp=\wp(z;0,1)$ is a solution of the differential equations \cite{Kamke77,Polyanin2018} 
\BEQ
\frac{\D^2 \wp(z)}{\D z^2} + 6 \wp^2(z) = 0 \;\; ; \;\;
\left( \frac{\D \wp(z)}{\D z}\right)^2 + 4 \wp^3(z) + 1=0
\EEQ
and we use the notation $\wp=\wp(z;g_2,g_3)$ in terms of the invariants
\BEQ
g_2 = 60 {\sum_{N,M}}' \frac{1}{\bigl( 2 M \omega + 2 N \omega'\bigl)^4} \;\; , \;\;
g_3 = 140 {\sum_{N,M}}' \frac{1}{\bigl( 2 M \omega + 2 N \omega'\bigl)^6}
\EEQ
where the sums over all pairs of integers $(M,N)$ exclude the pair $(0,0)$. The discriminant $\Delta = g_2^3-27 g_3^2=-27<0$ in the case at hand. 
The $\wp$-function belongs to the most simple elliptic functions since it has a single p\^ole of second order $\wp(z) \sim z^{-2}$ in the fundamental period parallelogram $0\, 2\omega\, 2\omega+2\omega'\ 2\omega'$
which for $\Delta<0$  has the form of a rhombus. 
One has the identities $\wp(z^*)=\wp(z)^*$ and  $\wp(z;g_2,-g_3)=-\wp(\II z;g_2;g_3)$ \cite[(18.2.16)]{Abra65} and the scaling 
$\wp\bigl(t z;t^{-4}g_2,t^{-6} g_3\bigr)= t^{-2} \wp(z;g_2,g_3)$ \cite[(18.2.13)]{Abra65}. Hence, along the imaginary axis, $\wp(\II y;0,1)$ is a real-valued and symmetric function of $y$, with
period $2\sqrt{3} \omega_2\simeq 5.299916$ since $\omega_2=\frac{\Gamma(1/3)^3}{4\pi}\simeq 1.529954$ in the equi-anharmonic case, defined by $g_2=0$ and $g_3=1$ \cite{Abra65}. In this case, the two periods
are $2\omega, 2\omega'=\omega_2 \bigl( 1 \mp \II \sqrt{3}\bigr)$. Finally, 
as an elliptic function of order 2, $\wp(\II y;0,1)$ has exactly two zeros in the fundamental period parallelogram, which occur at 
\BEA
y_{0}^{(\pm,N)} = \pm \frac{2\omega_2}{\sqrt{3\,}} +N 2\sqrt{3\,}\omega_2 \simeq \left\{ 
\begin{array}{l}  1.176639 + N 5.299916 \\ 3.533277 + N 5.299916 \end{array} \right.
\EEA
with $N\in\mathbb{Z}$. 

\appsection{B}{Fractional calculus}
A brief summary of results of fractional calculus is given, see e.g. \cite{herrmann2011fractional,Podlubny99,Diethelm2010} for further details. 
As a motivation, consider Abel's integral equation
%
 \begin{equation}
g(x) =\int_b^{z}\frac{f(y)\,\D y}{(x-y)^\eta}, \quad 0<\eta<1, 
 \end{equation}
where $g(x)$ is assumed known. Laplace transformations shows that the requested $f(x)$ is given by
\begin{equation} \label{B2}
f(x)=\frac{\sin\pi\eta}{\pi}\frac{\D}{\D x}\int_b^x \frac{g(y)\,\D y}{(x-y)^{1-\eta}},\quad 0<\eta<1.
 \end{equation}
 Without loss of generality, let from now on $b=0$. The fractional Caputo derivative of order $\eta$, written as ${^c}D^\eta f(x)$, is defined as
 \begin{equation} \label{Dc}
{^c}D^\eta f(x):=\frac{1}{\Gamma(1-\eta)}\int_0^x\frac{f'(y) \D y}{(x-y)^\eta}.
 \end{equation}
where $f'(x)$ denotes the standard first-order derivative. The Caputo derivative is distinguished from the  fractional order Riemann Liouville derivative, defined as
\begin{equation} \label{Dr}
^{r}D^\eta f(x):=\frac{1}{\Gamma(1-\eta)}\frac{\D}{\D x}\int_0^x\frac{f(y)\,\D y}{(x-y)^\eta}
 \end{equation}
where again $0<\eta<1$. The solution (\ref{B2}) can be written as $f(x)=\frac{\sin\bigl(\pi \eta\bigr)\Gamma(1-\eta)}{\pi}\, {^r}D^{\eta} g(x)$. 
Using the Leibniz rule for differentiation, it follows that the two definitions (\ref{Dc},\ref{Dr}) of fractional derivatives are related
\begin{equation}
^{r}D^\eta f(x)={^c}D^\eta f(x)+\frac{f(0)}{\Gamma(1-\eta)x^\eta}.
 \end{equation}
In particular, the Caputo derivative of a power function $f(x)=x^\beta$ (with $\beta\ne 0$) is 
\begin{equation}\label{power}
^{c}D^\eta x^\beta=\frac{\Gamma(\beta+1)}{\Gamma(\beta+1-\eta)}x^{\beta-\eta},\quad\beta\neq0
 \end{equation}
whereas for a constant function (i.e. $\beta=0$), the Caputo definition (\ref{Dc}) leads to the result 
\begin{equation}\label{constant}
^{c}D^\eta \,1=0.
 \end{equation}
familiar for a standard derivative. On the other hand, the Riemann-Liouville definition 
(\ref{Dr}) rather gives
\begin{equation}
^{r}D^\eta \,1=\frac{1}{\Gamma(1-\eta)}x^{-\eta}.
 \end{equation}
 which renders the formulation of initial-value problem in a fractional differential equation difficult. 
For a linear combination of functions
$f(x)=a_1f_1(x)+f_2(x)$, both definitions of the fractional derivative satisfy the linear relation:
\begin{equation}\label{power2}
D^\eta f(x)=a_1 D^\eta f_1(x)+a_2D^\eta f_2(x).
 \end{equation}
However, neither of them satisfy the semi-group property for iterative  derivatives, i.e.
\begin{equation}\label{power3}
D^{\eta_1} D^{\eta_2} f(x)\neq D^{\eta_1+\eta_2} f(x).\qquad \eta_1, \eta_2\in (0,1).
 \end{equation}
The property eq.~(\ref{constant}) motivates our choice of the Caputo fractional derivative in this work. It appears to us as the most suitable to describe the physical scenario of interest. The analysis of fractional calculus, within the context of diffusion equations (for either Riemann-Liouville or Caputo derivatives) is  studied in detail in \cite{Diethelm2010}. 


\end{document}